\UseRawInputEncoding
\documentclass[prl,aps,floats,superscriptaddress,floatfix,twocolumn]{revtex4}
\usepackage{amssymb,amsmath}
\usepackage{amsmath,amssymb}
\usepackage{graphicx}
\usepackage{psfrag}
\usepackage{color}
\usepackage{dcolumn}
\usepackage{bm}
\usepackage{soul}
\usepackage[normalem]{ulem}

\def\beq{\begin{equation}}
\def\eeq{\end{equation}}
\def\bea{\begin{eqnarray}}
\def\eea{\end{eqnarray}}
\begin{document}
%%%%%%%%%%%%%%%%%%%%%%%%%%%%%%%%%%%%%%%%%%%%%%%%%%%%%%%%%%%%%%%%%%%%%%%%%%%%%%%%
\title{Stiffening or softening of elastic media: Anomalous elasticity near phase transitions}
\author{Sudip Mukherjee}\email{sudip.bat@gmail.com}
\affiliation{Barasat Government College,
10, KNC Road, Gupta Colony, Barasat, Kolkata 700124,
West Bengal, India}
\affiliation{Condensed Matter Physics Division, Saha Institute of
Nuclear Physics, Calcutta 700064, West Bengal, India}
\author{Abhik Basu}\email{abhik.123@gmail.com, abhik.basu@saha.ac.in}
\affiliation{Theory Division, Saha Institute of
Nuclear Physics, 1/AF Bidhannagar, Calcutta 700064, West Bengal, India}
\date{\today}

\begin{abstract}
We present the general theory of Ising transitions in  isotropic elastic media with vanishing thermal expansion. By constructing a minimal model with appropriate spin-lattice couplings, we show that in two dimensions near a continuous transition 
the elasticity is anomalous in unusual ways: the system either significantly stiffens with a hitherto unknown unique positional order logarithmically stronger than quasi-long range 
order, or, as the inversion-asymmetry of the order parameter in its coupling with strain increases, it destabilizes when system size $L$ exceeds a finite threshold. 
At three dimensions, stronger inversion-asymmetric couplings induce instability to the long-range positional order for all $L$. Sufficiently strong order parameter-displacement couplings can also turn the 
phase transition  first order at all dimensions, concomitant with finite jumps in the elastic modulii across the transition.  %Our work paves the way to generic theories on zero thermal expansion systems.
 Our theory establishes a {\em one-to-one correspondence}
between the order of the phase transitions and anomalous elasticity near the transitions.
 
 %the elastic modulii in a system of size $L$ diverges with the variance of the local displacement fluctuations scaling as $[\ln (L/a_0)]^{2/3}$ and the  displacement correlation function scaling as $[\ln (r/a_0)]^{2/3}$ for weak inversion-asymmetric couplings between the local displacement-order parameter; $a_0$ is a small-scale cutoff and $r$ is a separation. %,  giving a novel state of the elastic media. 
 %The elastic constants can also vanish for $L$ exceeding a non-universal value, making the system unstable, with sufficiently strong inversion-asymmetry. These couplings can turn the phase transition  first order as well, when the elastic modulii do not diverge, but show jumps proportional to the jump in the order parameter, across the transition. For a bulk system, the elastic modulii do not diverge for weak asymmetry, but can vanish across a second order transition giving instability for strong asymmetry, or display jumps across a first order transition. This theory applies to compressible systems with vanishing strain in the zero-stress states as well. %These can be controlled by the coupling constants for the preferential couplings of the local displacements with a particular state of the order parameter. 
 %{\em In-vitro} experiments on binary fluids embedded in a polymerized network, magnetic colloidal crystals or magnetic crystals could test these predictions.
\end{abstract}
\maketitle

 %Phase transitions are a central question in equilibrium statistical mechanics. These can be first order with a finite jump of the order parameter, or second order with the order parameter vanishing smoothly as the transition temperature is approached. By now the general theory of phase transitions are well-developed and have been successfully applied over a range of physical systems~\cite{stanley,chaikin}. 

 Elastic media, e.g., crystals, are paradigms of {\em broken continuous symmetry phases} with positional order (PO) in condensed matter systems.  The interplay between the broken symmetry Goldstones modes (i.e., phonons) and the long-range correlated order parameter close to continuous phase transitions in broken symmetry phases can significantly affect the material properties of systems, e.g., 
  binary alloy crystals, polymerized or tethered networks in binary fluids~\cite{style-prx,nat-phys}, magnetic crystals and even {\em in-vivo} systems like biological cells~\cite{membrane-less}. 
  Previous  studies~\cite{moura,john-superfluid} suggested that the universal scaling properties of the second order transitions are unaffected by the elasticity, although the elastic modulii could get depressed and display anomaly at the onset of supersolid transition~\cite{john-superfluid} as a result of the elasticity-superfluid order coupling. In a seminal study, Ref.~\cite{berg-halp} discussed the generic instability of an elastic solid near a continuous Ising transition, except when $dT_c/dV=0$, where $T_c$ and $V$ are the Ising transition temperature, and system volume, respectively. In this case, Ref.~\cite{berg-halp} argued that the spin and the elastic degrees of freedom decouple,  leaving each one unaffected by the other, with vanishing or zero thermal expansion (ZTE) in the thermodynamic limit.  %the position and Ising spin fluctuations, respectively, are statistically identical to those in an isolated crystal, and in  an Ising system on a rigid lattice in the long wavelength
   %limit.
 
 %is
 %\begin{equation}
 % $\Delta_u\equiv \langle {\bf u}({\bf x})^2\rangle \sim \frac{T}{ a}\ln \left(\frac{L}{a_0}\right)$, at temperature $T$, implying quasi-long range order (QLRO), where $a$ is a elastic modulus. At higher dimensions D$>$2, $\Delta_u$ is a constant, independent of $L$, corresponding to long-range order (LRO).
 %\end{equation}

 %The statistical mechanics of elastic media has generated considerable among theoretical and experimental physicists. 

% [more introduction to be written]

  In this Letter, we specifically study continuous  Ising transitions in isotropic elastic media, e.g., a gel, with ZTE and investigate the measurable macroscopic properties near the phase transition. To this end, we present the general theory of Ising transitions in isotropic elastic media with ZTE, by constructing a suitable minimal model that {\em includes} generic spin-lattice interactions while maintaining the ZTE conditions.  %Given the potential applications of ZTE materials, a generic theory of phase transitions in a minimal model of ZTE would be very welcome. 
 Unexpectedly in this model, we find anomalous elasticity,  belonging to a {\em hitherto unstudied universality class}, making two-dimensional (2D) elastic media either {\em stiffer} or {\em unstable} near the transitions.  This shows that in an elastic solid coupled to Ising spins, the position and spin fluctuations are not independent of each other in the ZTE limit, or $dT_c/dV=0$.  Our theory revises the conclusions of Ref.~\cite{berg-halp} for an isotropic elastic media at $dT_c/dV=0$. 
 
  For our work, we conceptualize a schematic, purpose-built minimal spin-lattice model with interactions near the ZTE limit. We consider  a conceptual spring network model for gels, having each node carrying an Ising spin interacting with its nearest neighbor ferromagnetically, to illustrate the generic macroscopic properties near the phase transitions.  We delineate the interactions from the physical considerations that both the spin-spin exchange interaction strengths and the spring constants should be locally affected by the interactions, which are chosen to maintain ZTE. For simplicity, we assume that the spin-spin interaction rises quadratically with the local strains for small strains. We further allow a local spin-lattice coupling that depends quadratically on the strain but linearly on the spin, and provides  ``local magnetic field-like'' contributions. This manifestly breaks the Ising symmetry, being {\em asymmetric} or {\em selective} on the local spin state. Although such an Ising symmetry breaking term is unexpected in a ZTE system with magnetic or electronic origin, it however can in-principle exist in potential soft-matter realizations of the system~\cite{odd-phi}.  Quadratic dependence on strains of the interactions ensures vanishing thermal average of the strain, ensuring ZTE. Alternatively, we could imagine the spring constants of the gel to depend both linearly and bilinearly on the nearest neighbor spins. Instead of studying the discrete lattice model directly, we take
  a Landau-Ginzburg (LG) continuum approach for this system, that is particularly suitable to extract large-scale universal properties~\cite{chaikin}.  In this approach, we describe the system by a continuous Ising order parameter $\phi$ and symmetric strain tensor $u_{ij}\equiv (\partial_i u_j + \partial_j u_i + \partial_i u_m \partial_j u_m)/2$~\cite{chaikin}, where $\bf u({\bf x})$ is the displacement of position $\bf x$ in the undistorted system~\cite{chaikin,landau}.  We develop a generic and testable theory of the system as a thin, 2D sheet and in the bulk (dimension $d>2$). %Focusing on a system without any intrinsic internal strain, 
 Our most surprising result is that
   at 2D close to a second order transition at temperature $T=T_c$, the elasticity is {\em anomalous}: The system either significantly {\em stiffens} or {\em softens}, controlled by the order parameter-strain couplings.     
    When the order parameter-strain couplings are {\em weakly asymmetric} in $\phi$, the elastic modulii diverge in the long wavelength limit, and the system considerably stiffens {\em vis-a-vis} away from $T_c$. Indeed, we find for a system of size $L$ and microscopic cutoff (of the order of lattice constants) $a_0$%the displacement fluctuation variance %The  variance $\Delta^T_u$ of the local displacement fluctuations close to a second order transition is {\em significantly suppressed} vis-a-vis away from $T_c$. We find the variances 
 \begin{equation}
  \Delta^a_u\equiv \langle {\bf u}^a({\bf x})^2\rangle \propto [\ln (L/a_0)]^{2/3},\;a=T,L\label{eq1}
  %=\frac{T_c}{4\pi \mu_R} [\ln (L/a_0)]^{2/3}\label{eq1}
 \end{equation}
 near $T_c$, which although grows indefinitely, it does so considerably slower than the usual $\ln (L/a_0)$-dependence on $L$ in a 2D crystal without any transition. Here, ${\bf u}^a({\bf x})$ is
    the inverse Fourier transform of ${\bf u}^a({\bf q})$, the component of ${\bf u}({\bf q})$ transverse ($a=T$) and longitudinal ($a=L$) to the wavevector $\bf q$; ${\bf u}^T({\bf x})\cdot {\bf u}^L({\bf x})=0$.  
   Further, the correlation function %; $\mu_R$ is the amplitude of the renormalized version of bare or unrenormalized shear modulus $\mu$.
 %This, although still diverges for $L\rightarrow\infty$, corresponds to an order far stronger than the well-known QLRO,  observed 
%away from the phase transition, or in a single-component elastic medium. 
\begin{equation}
 \langle [{\bf u}^a({\bf x})- {\bf u}^a({\bf x'})]^2\rangle %\approx \frac{T_c}{4\pi \mu_R} 
 \propto [\ln (r/a_0)]^{2/3}, \label{corr-func}
\end{equation}
for large $r\equiv |{\bf x-x}'|$ near $T_c$,
that is significantly slower growing with $r$ than the more conventional $\ln(r/a_0)$ scaling  in an ordinary 2D crystal. Equations~(\ref{eq1}) and (\ref{corr-func}) illustrate a novel positional order {\em logarithmically} stronger than the conventional quasi-long range order (QLRO), that we name {\em SQLRO}, and constitute an entirely { new, heretofore unstudied universality class} for these systems.  This is {\em not} the only state of the system  near $T_c$. 
Increasing degree of the inversion-asymmetry of $\phi$ in its coupling with $u_{ij}$
  destabilizes the system with only short range order (SRO)  as soon as the system size $L$ exceeds a finite threshold. The order parameter-strain couplings can turn the phase transition, that is otherwise second order, into a first order one. In that case the elastic modulii are anomalous though in a different way: They do not diverge;
  instead, they and hence $\Delta^{a}_u$ (which now shows QLRO) display {\em finite jumps}  across the transition. %Stronger asymmetry can trigger instability as well. 
  
 For bulk three-dimensional (3D) samples, the elasticity is non-anomalous for weak inversion-asymmetry; the system displays positional long range order (LRO)  near $T_c$, indistinguishable from its behavior without any transition. Stronger asymmetry can make the elastic modulii vanish for all $L$, and hence destabilizes the system for any $L$, unlike 2D. Similar to 2D,  $\phi-u_{ij}$-couplings can turn the phase transition a first order, across which the elastic modulii and hence $\Delta^a_u$ display  finite jumps. %Again stronger asymmetry distablizes the system. 

%   \begin{figure}[htb]
%  \includegraphics[width=5cm]{phase-diag-L-overline-g1.pdf}
%  \caption{Schematic phase diagram in the $\overline g_1^2 - L$ plane in 2D with second order transition. The red curved line demarcates regions with positional order (PO) and no order or short range order (SRO). The region left to the vertical broken blue line corresponds to systems with arbitrarily large $L$ retaining PO. The region between the vertical blue line and the curved red line corresponds to systems having a finite $L<L_m$, a threshold value maintaining PO; for $L>L_m$ only SRO is possible (see text). }\label{fig2}
% \end{figure}

 %\begin{figure}[htb]
 % \includegraphics[width=5cm]{phase-diag1-g1-overline-g1.pdf}
 % \caption{Schematic phase diagram in the $\overline g_1/sqrt\mu - g_1$ plane. The middle light green region is where $\mu>0$ corresponding to positional order - SQLRO with second order transition and QLRO with first order transition in 2D and LRO in 3D. The region outside has $\mu<0$ implying loss of positional order or short range order (see text).}\label{fig3}
 %\end{figure}
 
 We now outline the derivation of these results; more details and additional results are available in the associated long paper (ALP)~\cite{alp}. 
 
 The LG free energy functional $\cal F$ of our minimal model, obtained phenomenologically by gradient expansions of the fields,  must be rotationally and translationally invariant~\cite{chaikin,landau}. Hence, its dependence on $\bf u(x)$ must be through $u_{ij}$. The form of $\cal F$, after dropping irrelevant terms, valid  near a second order transition of $\phi$, and for length scales much larger than $a_0$ is %The system is described by a scalar order parameter field $\phi ({\bf r})$ and a $d$-dimensional displacement field $\bf u ({\bf r})$ giving local displacements~\cite{landau,chaikin}. 
 %We use general symmetry considerations to write down the free energy functional $\cal F$:
 \begin{eqnarray}
 {\cal F} &=& \int d^dx \Big[\frac{\tilde r}{2}\phi^2 + \frac{1}{2}({\boldsymbol\nabla}\phi)^2 + v\phi^4 + \mu (\nabla_i u^T_j)^2 + \frac{\tilde\lambda}{2} (\nabla_i u^L_j)^2 \nonumber \\
 &&+ \left(g_1\phi^2 + \overline g_1 \phi\right) (\nabla_i u^T_j)^2 + \left(g_2\phi^2 + \overline g_2 \phi\right) (\nabla_i u^L_j)^2\Big],\label{free2}
\end{eqnarray}
%where% $\mu$ and $\tilde \lambda$ are the unrenormalized shear and bulk elastic modulii, respectively, %$u_i^L({\bf x})$ and $u_i^T({\bf x})$ are the %inverse Fourier transforms of $u_i^L({\bf q})$ and $u_i^T({\bf q})$; $\tilde\lambda=\lambda+2\mu$,
%transverse and longitudinal parts of ${\bf u}({\bf x})$, 
%$g_1,g_2,\overline g_1,\overline g_2$ define the anharmonic order parameter-strain coupling. Clearly, 
 Here, $\tilde r \equiv T-T_c$, $v>0$ for thermodynamic stability. The first three terms on the rhs of (\ref{free2}), together identical to ${\cal F} ({\bf u}=0)$, constitute the LG free energy for the Ising model near its critical point in a rigid uniform lattice in the absence of any external field~\cite{chaikin}. The $\mu$- and $\tilde\lambda$-terms give the elastic free energy of a crystal with $\mu,\,\tilde\lambda$ being the shear and bulk elastic modulii~\cite{chaikin}. There is no bilinear term in ${\bf u}^L,\,{\bf u}^T$ due to these being mutually orthogonal. 
 Each of the mixed anhamornic terms in (\ref{free2}) carries simple physical interpretations consistent with the interactions in the schematic spin-lattice microscopic model above. For instance, $\phi$-dependent corrections to the local elastic modulii are given by $\mu(\phi)=\mu + g_{1}\phi^2 + \overline g_{1}\phi,\, \tilde\lambda(\phi)=\lambda + 2g_{2}\phi^2 + 2\overline g_{2}\phi$. Alternatively, we could define a local critical temperature $T_c^* = T_c - 2  \left[g_1(u_{ij})^2 + g_{20}(u_{ii})^2\right]$, and a term formally analogous to ``local magnetic field'' $h_\phi = - \left[\overline g_1(u_{ij})^2 + \overline g_{20} (u_{ii})^2\right]$.  Couplings $g_1,\,g_2\geq 0$ due to thermodynamic stability reasons; $\overline g_1,\,\overline g_2$, the ``asymmetry'' parameters, have arbitrary signs, and violate the Ising symmetry of $\phi$.  Such couplings, which are odd in $\phi$, are frequently used in mixed soft matter systems, where $\phi$ represents the local concentration difference in the two components~\cite{odd-phi},  although these odd in $\phi$-anharmonic terms vanish in systems with Ising symmetry at the microscopic level, e.g., in a magnetic crystal. In order to generalize the scope of our work, we include the Ising symmetry-breaking soin-lattice interactions, and study the model.  The form of $\cal F$ applies to systems with vanishing strain in the zero-stress states~\cite{foot3}:  $\langle u_{ij}\rangle =0$ identically in the absence of any external stress implying vanishing thermal expansion.   Furthermore, (\ref{free2}) implies $dT_c/dV=0$~\cite{alp}.  The $\phi-u_{ij}$ anhamornic terms in (\ref{free2}) are not considered in Ref.~\cite{berg-halp} even when $dT_c/dV=0$.

\begin{figure}[htb]
 \includegraphics[width=4.2cm]{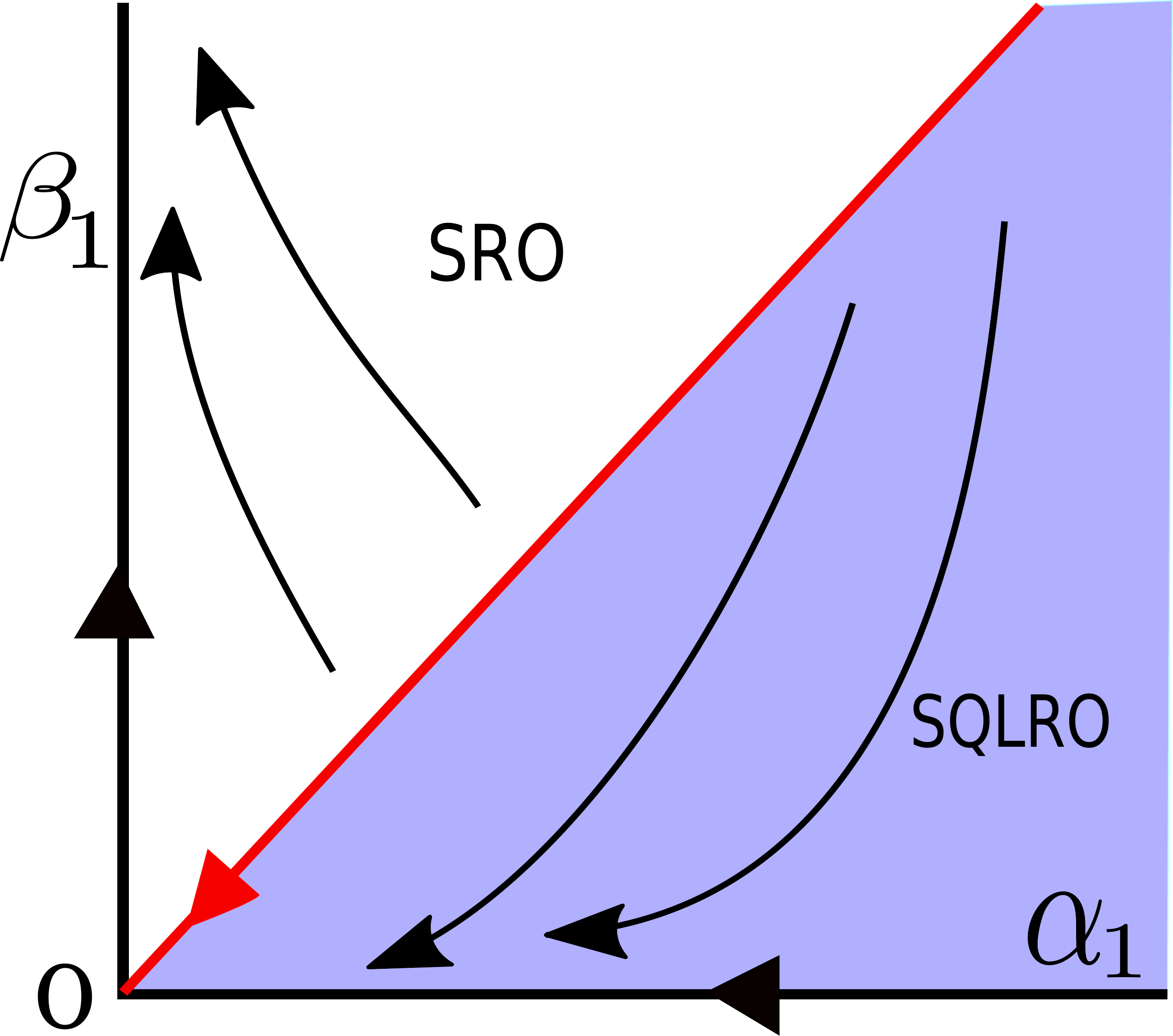}\hfill \includegraphics[width=3.7cm]{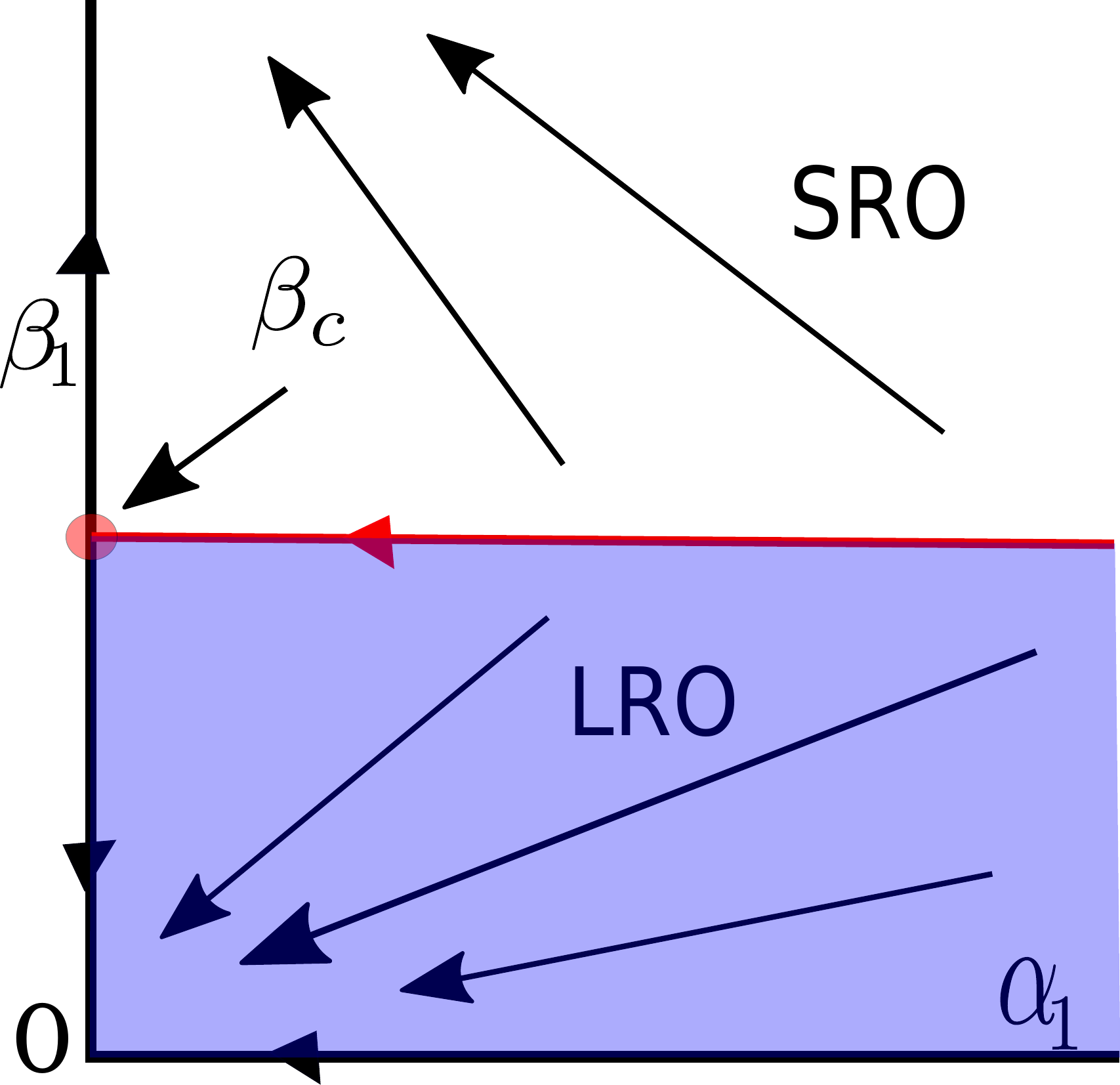}
 \caption{ (color online) RG flow diagram in the $\alpha_1-\beta_1$ plane. (left) In 2D; the inclined red line is the separatrix given by Eq.~(\ref{sep-eq}); (right) $d>2$:  The horizontal red line is the separatrix given by $\beta=\epsilon/2$. The small circle on the $\beta_1$-axis is the unstable fixed point $(0,\,\beta_c)$. Other arrows indicate the flow directions (see text).}\label{rg-flow}
\end{figure}

In the absence of the anharmonic effects, $\cal F$ implies %that the variances %of the displacement follow
\begin{equation}
 \langle {\bf u}^T({\bf x})^2\rangle = \frac{T}{2\mu}\ln \left(\frac{L}{a_0}\right),\,\langle {\bf u}^L({\bf x})^2\rangle = \frac{T}{\tilde\lambda}\ln \left(\frac{L}{a_0}\right) \label{gauss-scal}
\end{equation}
at all temperatures $T$. Near $T_c$, $\phi$-fluctuations are scale-invariant, and thus can ``connect'' distant parts of the system, creating a correlated background for the elastic deformations. The combination of thermal fluctuations and anhamornic effects can substantially modify the scaling behaviors  given in (\ref{gauss-scal}). To study this, we perform a one-loop momentum-space renormalization group (RG) analysis~\cite{chaikin} of the model (\ref{free2}).  Dimensional analysis gives that $v$ has a critical dimension of 4, whereas 2 is that of $g_1,\,g_2,\,\overline g_1,\,\overline g_2$~\cite{alp}; see also~\cite{john-superfluid}. Therefore, the phase transition of $\phi$ is unaffected by (assumed small)  bare or unrenormalized $g_1,\,g_2,\,\overline g_1,\,\overline g_2$.   The RG calculation is performed by integrating over the short wavelength Fourier modes of $\phi({\bf x})$ and ${\bf u}({\bf x})$, followed by rescaling of lengths, and the long wavelength parts of $\phi$ and $\bf u$~\cite{chaikin,ma}.   From the structure of (\ref{free2}),  ${\bf u}^T$ and ${\bf u}^L$ are mutually independent. We discuss the calculation for the renormalized correlation function of ${\bf u}^T$; the same for ${\bf u}^L$ is done analogously.  The perturbative RG procedure outlined above gives the following differential recursion relations for 
$\mu,\,g_1,\,\overline g_1$.  %: {\sethlcolor{cyan}\hl{Here we do not defined what is $\Lambda$, which is actually define later in the highligted text in same color}}
%the %model parameters $\mu,\,g_1,\overline g_1$:
\begin{eqnarray}
 \frac{d\mu}{dl}&=&%T_cg_1\frac{S_D\Lambda^\epsilon}{(2\pi)^D}-\frac{T_c\overline g_1^2}{2\mu}\frac{S_D\Lambda^\epsilon}{(2\pi)^D},\\
 \mu\alpha_1 - \mu\frac{\beta_1}{2},\label{mu-flow}\\
 % \frac{d\alpha_1}{dl}&=&-\epsilon\alpha_1-3\alpha_1^2-\frac{\beta_1^2}{8}+\frac{\alpha_1\beta_1}{2},\label{al1}\\
 %\frac{d\beta_1}{dl}&=&-\epsilon\beta_1 +2\beta_1^2+6\alpha_1\beta_1.\label{be1}
\frac{dg_1}{dl}&=&-\epsilon g_1- 2\frac{T_cg_1^2}{\mu}\frac{S_d\Lambda^\epsilon}{(2\pi)^d}-\frac{T_c\overline g_1^4}{8\mu^3}\frac{S_d\Lambda^\epsilon}{(2\pi)^d},\label{gflow}\\
\frac{d\overline g_1}{dl}&=&-\frac{\epsilon}{2}\overline g_1+\frac{T_c\overline g_1^3}{2\mu^2}\frac{S_d\Lambda^\epsilon}{(2\pi)^d}- \frac{2T_cg_1\overline g_1}{\mu}\frac{S_d\Lambda^\epsilon}{(2\pi)^d}. \label{gflow1}
 \end{eqnarray}
 Here, $S_d$ is the surface area of a $d$-dimensional unit sphere,
 $\epsilon\equiv d-2$, $\exp(l)$ is the length rescaling factor, and  $\Lambda=2\pi/a_0$ is an upper wavevector cut-off. %We define two effective dimensionless coupling constants
%\begin{equation}
% \alpha_1\equiv \frac{T_c g_1S_D}{(2\pi)^D\mu}\Lambda^\epsilon,\;\beta_1\equiv \frac{T_c \overline g_1^2 S_D}{(2\pi)^D \mu^2} \Lambda^\epsilon.
%\end{equation}
% The RG flow equations for $\alpha_1,\,\beta_1$ read
 Flow Eqs.~(\ref{mu-flow})-(\ref{be1}) in turn give the RG flow equations for two effective dimensionless couplings $\alpha_1\equiv \frac{T_c g_1S_d}{(2\pi)^d\mu}\Lambda^\epsilon,\;\beta_1\equiv \frac{T_c \overline g_1^2 S_d}{(2\pi)^d \mu^2} \Lambda^\epsilon$
\begin{eqnarray}
 \frac{d\alpha_1}{dl}&=&-\epsilon\alpha_1-3\alpha_1^2-\frac{\beta_1^2}{8}+\frac{\alpha_1\beta_1}{2},\label{al1}\\
 \frac{d\beta_1}{dl}&=&-\epsilon\beta_1 +2\beta_1^2+6\alpha_1\beta_1.\label{be1}
\end{eqnarray}
%We separately consider $d=2$ and $d>2$ below.
At 2D, $\epsilon=0$; the only fixed point of (\ref{al1}) and (\ref{be1}) is $\alpha_1=0,\,\beta_1=0$. Its stability property is intriguing: It is {\em stable} (i.e., attractive) along the $\alpha_1$-axis, but {\em unstable} (i.e., repulsive) along the $\beta_1$-axis.  Thus, starting from any initial value $(\alpha_1(l=0),\,\beta_1(l=0))=(\alpha_{10},\,0)$, the system flows {\em to} the origin (0,\,0) implying stability, whereas starting from any initial value $(0,\,\beta_{10})$ the system {\em flows away} from the origin, indicating instability. The separatrix,  an invariant manifold under RG in the $(\alpha_1,\,\beta_1)$ plane, that separates the stable phase from instability  can be obtained by using (\ref{al1}) and (\ref{be1}) along with the condition $d(\beta_1/\alpha_1)/dl=0$~\cite{john-tethered}. This, for small $\alpha_1,\,\beta_1$, is a straight line
\begin{equation}
 \beta_1=\Gamma_{1c}\alpha_1,\;\;\Gamma_{1c}\equiv\frac{1}{2}\left[-12+\sqrt{240}\right];\label{sep-eq}
\end{equation}
see Fig.~\ref{rg-flow} (left).  Separatrix (\ref{sep-eq}) is {\em repulsive}: by expanding around (\ref{sep-eq}), (\ref{al1}) and (\ref{be1})  give $\beta_1(l)/\alpha_1(l)\sim 1/l \rightarrow 0$, when $\alpha_{10},\,\beta_{10}$ lie in the stable region, i.e,  {\em below} the line (\ref{sep-eq}) in the $\alpha_1-\beta_1$ plane in the thermodynamic limit. Therefore,  systems whose starting parameters lie {\em below} the separatrix, not only flow to the origin, they also {\em move away} from the separatrix with increasing length scale; see Fig.~\ref{rg-flow} (left). Thus, in the long wavelength limit, the system is effectively identical with those having no inversion-asymmetry;  the inversion-symmetry of  $ \phi$, although absent at small scales, appears as an emergent symmetry in the thermodynamic limit.  On the stable side of the separatrix, using (\ref{mu-flow}) we find the renormalized, scale-dependent shear modulus $\mu(l)\sim l^{1/3}$, which implies wavevector-dependent $\mu(q) \approx\mu_R [\ln (\Lambda/q)]^{1/3}$, in the long wavelength limit, where $\mu_R$ is the amplitude of the renormalized shear modulus. %or $\mu(t)\approx [\ln t]^{1/3},\,t\equiv |T-T_c|/T_c$ as one approaches the critical point, 
 This means the elasticity is anomalous, analogous to the well-known anomalous elasticity in 3D smectics~\cite{smect}.
This in turn gives, as shown in detail in ALP~\cite{alp}
\begin{eqnarray}
 \langle |u^T({\bf q})|^2\rangle &\approx&\frac{T_c}{2\mu_R [\ln (\Lambda/q)]^{1/3}q^2},\label{uT1}
 %\langle [u_T({\bf x})-u_T({\bf x}')]^2\rangle &\approx&\frac{T_c}{2\mu_R}\left[\ln (r/a_0)\right]^{2/3},\label{uT2}
\end{eqnarray}
for small $q$ close to $T_c$. Inverse Fourier transform of (\ref{uT1}) gives (\ref{eq1}) and (\ref{corr-func}).
%in the long wavelength limit, i.e., large $r$ and small $q$. 
Thus, both $\langle |u^T({\bf q})|^2\rangle$ and $\langle [u^T({\bf x})-u^T({\bf x}')]^2\rangle$ are significantly suppressed in the long wavelength limit vis-a-vis  their values away from $T_c$, a hallmark of SQLRO. Results (\ref{uT1}) together with (\ref{corr-func}) illustrate the  novel state of the elastic media near  $T_c$, and define a { new universality class with unique features}  that has not been
studied before.  Thus, in a ZTE system with microscopic Ising symmetry, i.e., of magnetic or electronic origin, $\beta_1=0$ identically, and the results (\ref{uT1}) and (\ref{corr-func}) are {\em generic and observable} without any additional fine tuning of the control parameters. For ZTE systems of soft matter origin with non-zero microspic Ising symmetry breaking anharmonic spin-lattice interactions,  $\beta_1$ is non-zero microscopically. Thus to observe (\ref{uT1}) and (\ref{corr-func}) additional tuning making the ``RG initial values'' $\alpha_1(l=0),\,\beta_1(l=0)$ lying below the separatrix is required. 

%{{\sethlcolor{cyan}\hl{Here, $\Lambda=2\pi/a_0$ is an upper wavevector cut-off}}.

 We now turn to the region above the separatrix in Fig.~\ref{rg-flow} (left), that flows away from the origin. { This region is accessible in potential soft matter realizations of ZTE systems by additional tuning to make the ``RG initial values'' $\alpha_1(l=0),\,\beta_1(l=0)$ lie {\em above} the separatrix.} In this region, $\beta_1(l)$ diverges, whereas $\alpha_1(l)$ vanishes as $l$ exceeds a nonuniversal value of $l$, controlled by the microscopic model parameters. This in turn means that the system flows towards negative $\mu$. While it is not possible to follow these flows all the way to $\mu=0$ (since $\beta_1$ diverge there breaking down our perturbation theory), this signals breakdown of elasticity and loss of PO in large enough systems. This region of the parameter space therefore corresponds to a phase with only SRO. 

\begin{widetext}
  
\begin{figure}[htb]
  \includegraphics[width=4.5cm]{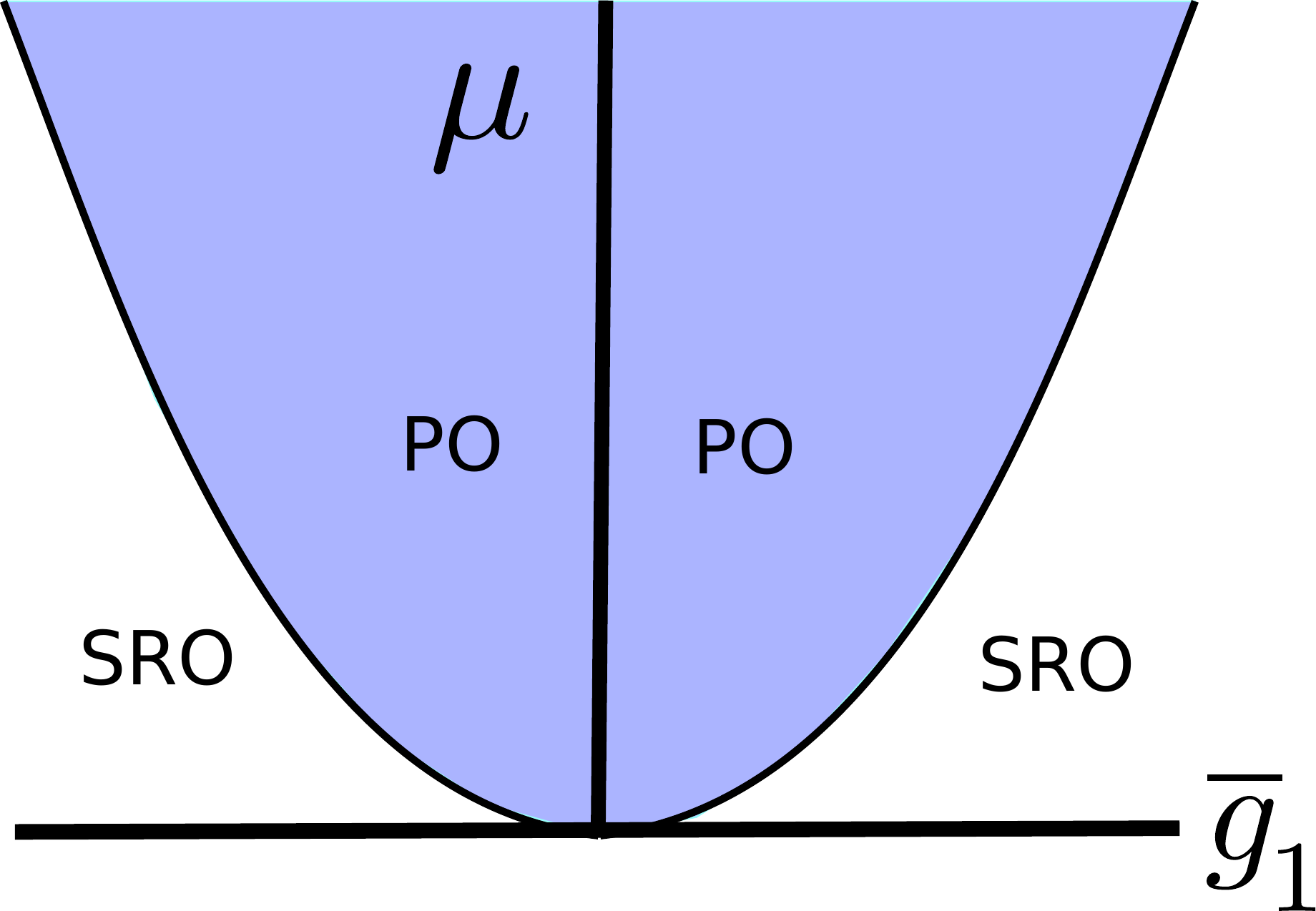}\hfill \includegraphics[width=3.7cm]{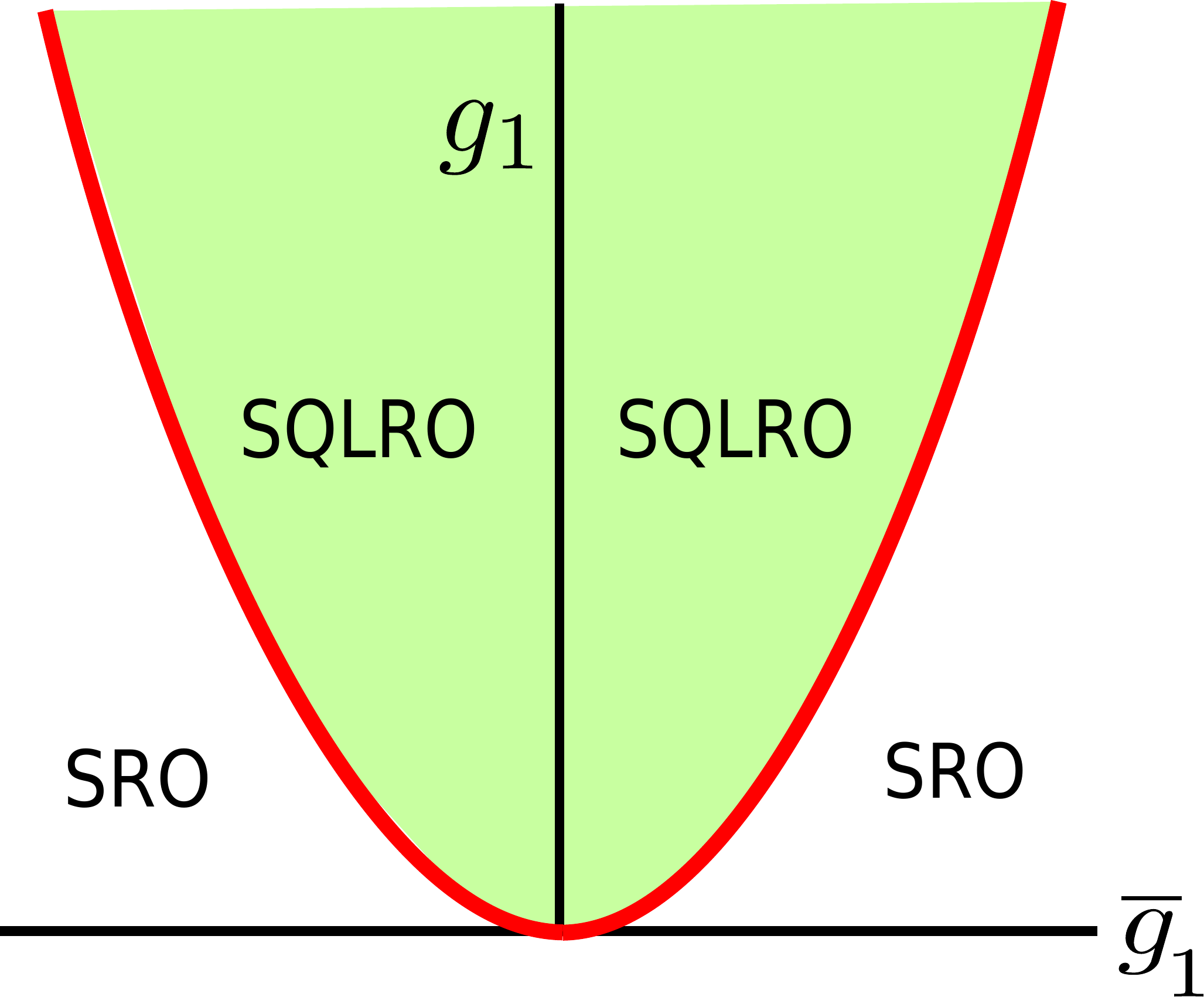}\hfill \includegraphics[width=4.1cm]{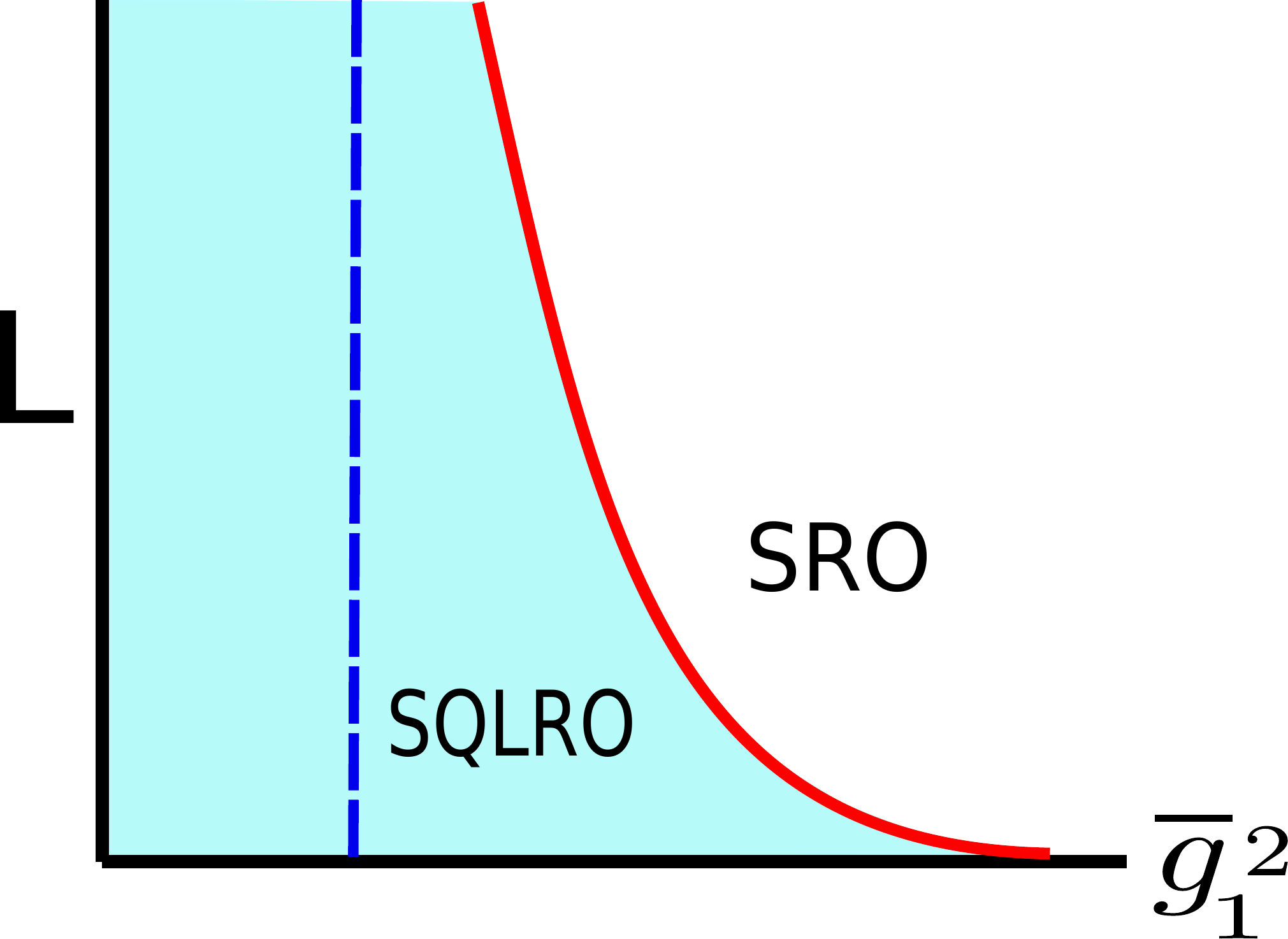} 
  \caption{ (color online) Schematic phase diagram in the (left) $\overline g_1 -\mu$ plane. The blue shaded region with $\mu>0$ corresponds to positional SQLRO  at 2D and LRO at 3D near $T_c$. The white region outside has SRO.  (middle)  $\overline g_1-g_1$ plane. The middle light green region indicates  SQLRO in 2D. The region outside has SRO.
   (right) $\overline g_1^2-L$ plane in 2D near $T_c$. The red curved line demarcates regions with SQLRO and SRO. The region left to the vertical broken blue line corresponds to SQLRO for any large $L$. The region between the vertical blue line and the curved red line corresponds to systems having a finite $L<\xi$, a threshold for PO; for $L>\xi$ only SRO is possible. See text.}\label{fig1}
 \end{figure}
 
  \end{widetext}

Whether or not the system destabilizes depends surprisingly upon the small-scale or unrenormalized values of the model parameters. The above results can be used to show that %the combination of the unrenormalized parameters
%\begin{equation}
 %$\Gamma_1\equiv
 %${\overline g_1^2}/{(\mu g_1)}= \Gamma_{1c}$ %,\label{basic1}
%\end{equation}
 %a {\em universal number} 
 %gives the instability threshold, i.e., 
 for $\Gamma_1 <(>)\Gamma_{1c}\equiv {\overline g_1^2}/{(\mu g_1)}$, a combination of the unrenormalized parameters,  SQLRO ensues (gets unstable with SRO). Phase diagrams in (i) $\overline g_1-\mu$  for a fixed $g_1$, and (ii) $\overline g_1-g_1$ for a fixed $\mu$ in Fig.~\ref{fig1} left and middle respectively follow directly from this threshold relation. % (\ref{basic1}).  
 Both phase boundaries are parabolas, as can be seen from the instability threshold.
 
% When $\mu$ is plotted against $\overline g_1$ for a fixed $g_1$ by using (\ref{basic1}), we get Fig.~\ref{fig1}(left). Instead, if $g_1$ is plotted against $\overline g_1$, Fig.~\ref{fig1}(right) is obtained.

On the unstable side of the separatrix, $\beta_1(l)\ll \alpha_1(l)$ for large $l$; (\ref{mu-flow}) reduces to
%\begin{equation}
 $\frac{d\mu}{dl}\approx -\mu \frac{\beta_1}{2}$.
%\end{equation}
Solving, we obtained
\begin{equation}
 \mu(l)\approx\mu(l=0)\left(\frac{1}{2\beta_{10}}\right)^{1/4}[l-\frac{1}{2\beta_{10}}]^{1/4},%\sqrt{\ln (L/a_0)-1/2},
 \label{mu-l}
\end{equation}
%{\sethlcolor{red}\hl{Why we use $|..|$ in the above expression}}\\
for large $l$. Of course, we cannot use (\ref{mu-l}) all the way to $\mu(l)\approx 0$ as $\beta_1(l)$ diverges and the RG scheme breaksdown. Instead, %we resort to the bare perturbation to draw an important conclusion on the physics on the unstable side of the separatrix. In fact 
as shown in ALP~\cite{alp}, we can conveniently use the bare perturbation theory to define a persistence or correlation length $\xi$ that is finite on the unstable side of the separatrix, given by $\mu(l=\ln(\xi/a_0))=0$. This gives, as shown in ALP~\cite{alp}
\begin{equation}
 \xi=a_0\exp\left[{2\pi\mu}/\{T_c\left({\overline g_{1}^2}/{2\mu} - g_{1}\right)\}\right],\label{xi}
\end{equation}
as one approaches the instability threshold from the unstable side~\cite{foot4}.
Thus, as $\overline g_1^2\rightarrow 2\mu g_1$, $\xi$ diverges as an {\em essential singularity}, surprisingly reminiscent of the behavior of the correlation length near the Kosterlitz-Thouless transition of the 2D XY 
model~\cite{chaikin,kt-paper}.  Physically, it is clear that as one moves from the stable region to the unstable region crossing the separatrix in the $\alpha_1-\beta_1$ plane, the system undergoes a structure phase transition from a phase with positional SQLRO to a phase with only SRO. Unsurprisingly, $\xi$ is infinite on the stable side of the separatrix. Equating $\xi$ with $L$, we get from (\ref{xi}) the maximum linear size that can show PO; see Fig.~\ref{fig1} (right) for a plot of $L$ versus $\overline g_1^2$.

%Setting $\mu(l=l^*)=0$, we get a critical system size $L^*=a_0\exp [1/(2\beta_{10})]$ beyond which any sense of positional order will be completely lost. A plot of $\mu(l)$ versus $L$ is shown is Fig.~\ref{fig3}.

%It is clear that as one moves from the stable region to the unstable region crossing the separatrix in the $\alpha_1-\beta_1$ plane, the system undergoes a structure phase transition from a phase with positional SQLRO to a liquid-like phase with only SRO.  This suggests that the structural phase transition is a second order transition~\cite{foot2}

% Analysis similar to the above gives results for $u^L_i$ that are exactly analogous to (\ref{corr-func}) and (\ref{uT1}), and a separatrix analogous to (\ref{sep-eq}) that separates stable region with positional SQLRO and instability with SRO; see ALP for details.

%. Below a separatrix analogous to (\ref{sep-eq}) in the plane of the effective coupling constants, the system stiffens with
%\begin{eqnarray}
% \langle |u_L({\bf q})|^2\rangle &=&\frac{T_c}{\tilde\lambda_R [\ln (\Lambda/q)]^{1/3}q^2},\label{uL1}\\
% \langle [u_L({\bf x})-u_T({\bf x}')]^2\rangle &=&\frac{T_c}{\tilde\lambda_R}\left[\ln (r/a_0)\right]^{2/3}.\label{uL2}
%\end{eqnarray}
%Here, $\tilde\lambda_R$ is the amplitude of the renormalized bulk modulus.
% In contrast, on the other side of the separatrix, there is instability whose analysis is identical that for $u_T$; see ALP for detail calculations.

We now consider bulk systems with $d>2$; $d=3$, i.e., $\epsilon=1$ is the physically relevant dimension. We focus on the fluctuations of ${\bf u}^T$, and an equivalent analysis for ${\bf u}^L$ can be done exactly in the same manner; see ALP~\cite{alp}. Flow equations (\ref{al1}) and (\ref{be1}) give two fixed points for $\epsilon>0$: (i) $(\alpha_1=0,\,\beta_1=0)$, i.e., the origin, which is linearly stable, and ($\alpha_1=0,\,\beta_1=\epsilon/2)$, unstable along the $\beta_1$-direction, but stable along the $\alpha_1$-direction; see Fig.~\ref{rg-flow} (right). In fact,  separatrix $\beta_1=\epsilon/2$, a straight line parallel to the $\alpha_1$-axis, determines that systems with microscopic model parameters in the region below it flow towards the origin rapidly. As shown in ALP, at this stable fixed point, $\mu$ does not diverge in the thermodynamic limit unlike in 2D, but is a constant. Thus elasticity is {\em non-anomalous},  with  $\langle {\bf u}^T({\bf x})^2\rangle$ being {\em bounded}, i.e., a constant independent of $L$, which is a telltale signature of positional LRO, and is indistinguishable from  its behavior away from $T_c$. On the other hand, systems whose starting parameters lie above the separatrix, flow towards the $\beta_1$-axis but away from the origin, making $\beta_1(l)$ diverge and $\alpha_1(l)$ vanish again rapidly. Correspondingly, $\mu(l)$ flows to zero for all $L$.  When $\beta_1$ increases beyond $\epsilon/2$, corresponding to  $\overline g_1$ rising above a threshold, the system undergoes a structural phase transition in which it loses LRO, and displays just SRO, akin to liquids. As shown in ALP~\cite{alp}, bare perturbation theory reveals that as soon as $\beta_{10}>2+2\alpha_{10}$, fluctuation-corrected $\mu$ becomes negative, giving complete loss of LRO, independent of $L$, unlike in 2D. Indeed, $\xi$ for which $\mu(\xi)=0$ abruptly changes from  very large to zero (or very small), as one crosses the separatrix from the stable to the unstable sides. See Fig.~\ref{fig1} (left) for a schematic phase diagrams in the $\overline g_1 -\mu$ plane for $d>2$.

%identifying $\xi$ with the positional correlation length $\xi_L$, we find that it drops from infinity (LRO) to vanishingly small (SRO) abruptly, suggesting a first order structural transition.

  %These results apply to any two-component incompressible elastic media. 
  For compressible  systems with bulk modulus $\tilde\lambda$ and longitudinal displacement ${ u}^L$, exactly analogous results and phase diagrams  for both 2D and 3D exist; see ALP~\cite{alp}.

%[nature of this phase transition]

%Proceeding exactly as above and shown in Ref.~\cite{alp}, we arrive at the exactly analogous results for $u_L$, {\em viz.}, for small enough $\overline g_2$, $\langle u_L({\bf q})^2\rangle$ is a constant independent of $L$ in the thermodynamic limit, corresponding to LRO. In contrast, as soon as $\overline g_2$ exceeds a nonuniversal threshold, $\langle u_L({\bf q})^2\rangle$ displays only SRO, implying again a structural phase transition.

{So far we have tacitly assumed that the second order phase transition of $\phi$ is unaffected by the displacement-order parameter couplings. {\em How correct is that?} We first consider the case with microscopic Ising symmetry, i.e., the couplings $\overline{g}_1,\,\overline g_2$ vanish. In this case, in order to have a second order transition, it is required that that under mode elimination, $v_e$, the fluctuation-corrected $v$ at any intermediate scale never  turns negative. This may not hold true for sufficiently strong order parameter-strain couplings. In the anticipation that $v_e$ can actually turn negative, we extend ${\cal F}$ by adding a $v_6\phi^6$-term in it with $v_6>0$ for thermodynamic stability reasons. We consider the inhomogeneous fluctuation corrections to $v$ that originate from $g_1,\overline g_1, g_2,\overline g_2$, that themselves do not depend upon $v$ explicitly. These contributions are {\em finite}, but {\em negative}: we get~\cite{alp}
\begin{eqnarray}
 \beta_c v_e&\equiv& \beta_c v-2dT_c^2\left(\frac{\beta_c^2 g_1^2}{4\mu^2}+\frac{\beta_c^2 g_2^2}{\tilde\lambda^2}\right)\frac{\Lambda^d}{(2\pi)^d},\label{v-tot}
\end{eqnarray}
valid for all $d \geq 2$. Now, for $v_e>0$, the $v_6\phi^6$-term is unnecessary. The phase transition of $\phi$ is unaffected by the order parameter-strain couplings, and remains a continuous transition belonging to the Ising universality class. %All our above conclusions (valid near $T_c$) hold good. 
If, however, $v_e<0$, then a $v_6\phi^6$-term must be taken into account for reasons of thermodynamic stability. In that case, $\phi$ now undergoes a first order transition with the order parameter $m\equiv \langle\phi\rangle$ jumping of magnitude $[|v_e|(2v_6)]^{1/2}$. We thus conclude that in ZTE systems with microscopic Ising symmetry, sufficiently strong spin-lattice couplings necessarily turn the second order transition into a first order one.

For non-zero inversion symmetry-breaking or the selectivity parameters, additional tuning is necessary to access a second order transition. This is because fluctuations generate a $g\phi^3$-term in $\cal F$,
$g$ is a coupling constant of arbitrary sign. This allows us to generalize the Landau-Ginzburg free energy $\cal F$ to $\tilde {\cal F}= {\cal F} +\int d^dx g_e\,\phi^3$, giving a generic liquid-gas like first order transition~\cite{chaikin,alp} with an order parameter jump $m=-g_e/(2 v)$ at a transition temperature $T^*=T_c+9g_e^2/(16 v)$~\cite{chaikin}. To proceed further, we integrate out the strains in $\tilde {\cal F}$ perturbatively in the coupling constants, generating a ``dressed''  free energy in terms of the dressed parameters, which depends only on $\phi$. By shifting $\phi\rightarrow \phi+\phi_0$ and choosing $\phi_0$ appropriately the effective coefficient of $\phi^3$ can be made to vanish in a manner analogous to the liquid-gas transition; see ALP for more details. In terms of the shifted $\phi$, $\tilde {\cal F}$ has the same form as $\cal F$, albeit with shifted model parameters. The requirement of ZTE remains satisfied due to the absence of terms in the free energy functional that are odd in the displacement $\bf u$. Note also that in this case $\beta v_e$ in (\ref{v-tot}) is to be supplemented by an additional finite  negative contribution $2dT_c^4 \left(\frac{\beta_c^4\overline  g_1^4}{32 \mu^4} + \frac{\beta_c^4\overline g_2^4}{\tilde\lambda^4}\right)\frac{\Lambda^d}{(2\pi)^d}$~\cite{alp}. Again, if $v_e>0$ a second order Ising transition follows, and a $v_6\phi^6$-term is not necessary. In the vicinity of this transition, our above results on the anomalous elasticity as above then ensue. On the other hand, if $v_e<0$ a first order transition ensues with an order parameter jump $ m= [|v_e|(2v_6)]^{1/2}$. Thus, even in the presence of Ising-symmetry breaking spin-lattice coupling terms, although the transition is generically first order, a second order Ising transition can be accessed by tuning the model parameters reminiscent of the second order transition in liquid-gas systems. This second order transition can get converted into a {\em different} first order one for sufficiently strong spin-lattice interactions. Across such first order transitions, the elastic modulii are finite, but still anomalous in the sense given below.}

Near a first order transition, fluctuations of $\phi$ do not have long range correlations, and as a result, all the corrections to $\mu$ at 2D are finite (and small). In a mean-field description that suffices near a first order transition we get, as shown in ALP
\begin{eqnarray}
\mu_{T<T^*}&=&\mu_{T>T^*} + (g_1 -{\overline g_1^2}/{\mu})m^2\neq \mu_{T>T^*},
\end{eqnarray}
valid at all dimensions, where $T^*$ is the first order transition temperature.  Thus, depending upon the relative magnitudes of $g_1$ and $\overline g_1$, $\mu(T<T^*)$ can be larger or small than $\mu (T>T^*)$, with a finite jump in its value at $T^*$, related to the jump in the order parameter. Therefore at 2D, $\langle {\bf u}^T({\bf x})^2\rangle$ should scale as $\ln (L/a_0)$ corresponding to QLRO with an amplitude that shows the jump. In contrast, across a second order transition, the elastic modulii do not display any jump; instead they either continuously increase or decrease (and approach zero for large enough systems) as $T_c$ is approached from either side.
At 3D, $\langle {\bf u}^T({\bf x})^2\rangle$ shows conventional LRO on both sides of $T^*$, differing only by a finite jump. At any dimension, for very large $\overline g_1$, $\mu(T<T^*)$ can even turn negative, signaling instability and SRO. A similar relation exists for $\tilde\lambda$.
We thus establish a {\em one-to-one correspondence} between the order of phase transitions and anomalous elasticity around the transition temperature. Detail derivation of all these results in addition to a
plethora of others  are available in ALP~\cite{alp}.%In particular, our theory reveals that a second (first) order phase transition of the order parameter is necessarily associated with a second (first) order structural transition.
%\begin{figure}[htb]
% \includegraphics[width=5cm]{3d-sep.pdf}
% \caption{RG flow diagram in the $\alpha_1-\beta_1$ plane in 3D. The blue line is the separatrix given by $\beta=\epsilon/2$. The small circle on the $\beta_1$-axis is the unstable fixed point $(0,\,\beta_c)$. Arrows indicate the flow directions (see text).}\label{rg-3d}
%\end{figure}

%All the results quoted above, together with a 
%summary: to be written later
We have thus developed the theory of Ising transitions in isotropic elastic media with vanishing thermal expansion. This theory predicts anomalous elasticity with the system either stiffening or softening near the transitions, being controlled by the strain-order parameter couplings.   These are in contrast to Ref.~\cite{berg-halp} due to the absence of the spin-lattice anharmonic interaction terms there even at $dT_c/dV=0$. Our theory should be a guideline to theoretically study of ZTE materials of diverse origin, including electronic magnetism and soft matter systems~\cite{exam}, which are of great demands in high precision applications~\cite{advsc,engg}.
  Experiments on purpose-built synthetic ZTE systems~\cite{advsc,engg,thermal,two-comp} in future should help to verify our results.   %We expect our theory to provide strong impetus towards  fu in this direction.

%in binary fluids in en elastic matrix, or on nearly incompressible bulk samples as well as by numerical simulations. We hope our work will stimulate future experimental and numerical studies of physically relevant systems and models.

{\em Acknowledgement:-} S.M. thanks 
the SERB, DST (India) for partial financial support through the TARE scheme [file no.: TAR/2021/000170]. AB thanks the SERB, DST (India) for
partial financial support through the MATRICS scheme
[file no.: MTR/2020/000406].


\begin{thebibliography}{99}
\bibitem{foot3} We did not include in a pure $\phi^3$-term in $\cal F$, as we assumed a generic continuous transition in the rigid limit of the lattice.
\bibitem{foot4} This condition of singularity from (\ref{xi}) is slightly different from (\ref{sep-eq}). This is because (\ref{xi}) is obtained by using the bare perturbation theory instead of the renormalized theory.%one that gives (\ref{basic1}).
\end{thebibliography}
\end{document}